\documentclass{article}
\usepackage{spconf,amsmath,graphicx}
\usepackage{booktabs}
\usepackage{subfig}
\usepackage{arydshln}
\usepackage{hyperref}
\usepackage{amsfonts}

\usepackage{bibspacing}
\title{Unifying Robustness and Fidelity: A Comprehensive Study of Pretrained Generative Methods for Speech Enhancement in Adverse Conditions}
\name{\begin{tabular}{c}Heming Wang$^{1*}$\thanks{*Work done during an internship at Tencent AI Lab.}, Meng Yu$^2$, Hao Zhang$^2$, Chunlei Zhang$^2$, Zhongweiyang Xu$^{4*}$, \\ Muqiao Yang$^{3*}$, Yixuan Zhang$^{1*}$, Dong Yu$^2$\end{tabular} }

\address{
$^1$The Ohio State University, USA\;\;\;\;\; $^2$Tencent AI Lab, USA \;\;\;\;\; \\$^3$Carnegie Mellon University, USA \;\;\;\;\;$^4$University of Illinois Urbana-Champaign, USA\\
}

\begin{document}
\ninept
\maketitle
\begin{abstract}
Enhancing speech signal quality in adverse acoustic environments is a persistent challenge in speech processing. Existing deep learning based enhancement methods often struggle to effectively remove background noise and reverberation in real-world scenarios, hampering listening experiences. To address these challenges, we propose a novel approach that uses pre-trained generative methods to resynthesize clean, anechoic speech from degraded inputs. This study leverages pre-trained vocoder or codec models to synthesize high-quality speech while enhancing robustness in challenging scenarios. Generative methods effectively handle information loss in speech signals, resulting in regenerated speech that has improved fidelity and reduced artifacts. By harnessing the capabilities of pre-trained models, we achieve faithful reproduction of the original speech in adverse conditions. Experimental evaluations on both simulated datasets and realistic samples demonstrate the effectiveness and robustness of our proposed methods. Especially by leveraging codec, we achieve superior subjective scores for both simulated and realistic recordings. The generated speech exhibits enhanced audio quality, reduced background noise, and reverberation. Our findings highlight the potential of pre-trained generative techniques in speech processing, particularly in scenarios where traditional methods falter. Demos are available at \href{https://whmrtm.github.io/SoundResynthesis}{https://whmrtm.github.io/SoundResynthesis}.

\end{abstract}
\begin{keywords}
speech enhancement, speech vocoder, speech codec, robustness, fidelity
\end{keywords}
\section{Introduction}
\label{sec:intro}

In real-world scenarios, speech signals are often degraded by background noise and room reverberation, leading to diminished clarity and comprehensibility. The main aim of speech enhancement is to mitigate the impact of such environmental disturbances.
The development of deep neural networks (DNN) has greatly advanced speech enhancement research. DNNs have shown remarkable proficiency in suppressing background noise and reverberation, yielding satisfactory enhancement results \cite{wang2018supervised}.
DNN-based enhancement techniques primarily focus on direct speech signal representations, aiming to establish mappings from noisy inputs to their corresponding clean targets.
These representations include but not limited to magnitude \cite{han2014tms,li2020online}, complex spectrograms \cite{choi2018phase,hu2020dccrn}, waveforms \cite{luo2018tasnet,pandey2021dense}, or a fusion of these features \cite{li2021two,wang2021nca} which are all intrinsically associated with the signals.
Despite the effectiveness of existing powerful enhancement baselines, their performance often notably deteriorates in real-world complicated scenarios.
The enhanced speech obtained by supervised learning based models in such challenging scenarios may retain strong noise or reverberation, and be accompanied by distortions and artifacts \cite{rao2021interspeech}.

To address these challenges, recent studies aim to leverage the potential of pre-trained models.
Some researchers utilized diffusion models to refine speech, employing them to regenerate clean speech based on enhanced priors acquired through pre-trained discriminative models \cite{wang2023cross,lemercier2023storm}.
Another avenue of investigation involves employing speech vocoders for speech resynthesis. 
For instance, VoiceFixer was proposed to address general speech restoration \cite{liu2022voicefixer}. It employs an enhancement model on mel-spectrograms and subsequently utilizes the HifiGAN \cite{kong2020hifigan} vocoder to resynthesize the clean speech.
Similarly, \cite{zhong2023audiomae} proposed to use masked autoencoders for speech restoration, and employs mel-to-mel mapping during pretraining to restore masked audio signals.
We believe that discrete representations stored in codebooks are more robust against various interference, and propose to employ speech codecs to perform speech enhancement.
The majority of existing research related to speech codecs \cite{borsos2023audiolm,wang2023valle} is primarily centered around text-to-speech tasks, relying heavily on text embeddings to ensure input stability.
Drawing inspiration from a parallel study in computer vision \cite{zhou2022codeformer}, which addresses blind face restoration through the regeneration of code tokens within a learned discrete codebook, we are motivated by its exceptional robustness against degradation in both synthetic and real-world datasets.
Furthermore, a relevant contribution by Wav2code \cite{hu2023wav2code} has also introduced the utilization of codebooks to enhance the resilience of speech representations. Notably, Wav2code focuses more on improving robust automatic speech recognition and operates on self-supervised learning (SSL) embeddings.

\begin{figure*}[t]
\includegraphics[width=0.75\linewidth]{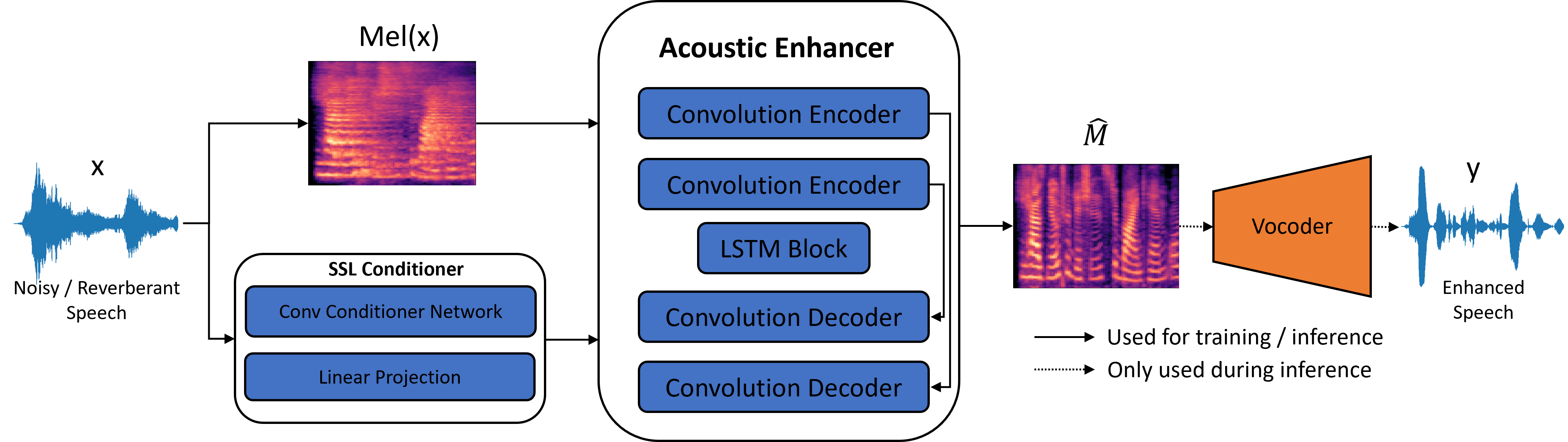}
\centering
\caption{The overview of the vocoder pipeline.}
\vspace{3mm}
\label{fig:vocoder}
\includegraphics[width=0.75\linewidth]{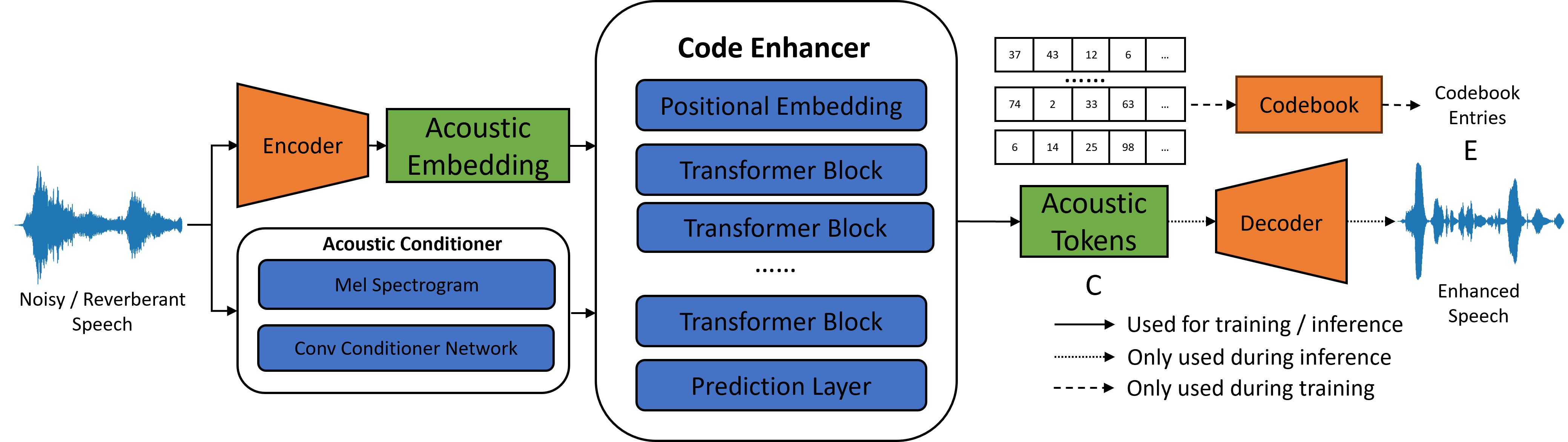}
\centering
\caption{The overview of the codec pipeline.}
\label{fig:codec}
\end{figure*}

This paper systematically investigates two pipelines: one based on a speech vocoder and the other on a speech codec. In both pipelines, our network processes a main input and an auxiliary input to enhance intermediate representations. These representations are then enhanced before generating the desired speech output. We choose this design because generative methods excel in addressing complex situations with significant information loss in speech signals. Utilizing pre-trained models also benefits by leveraging existing semantic or acoustic information, aiding in faithful resynthesis of the original speech and enhancing fidelity. Experimental outcomes with real and synthetic datasets demonstrate the superiority of our proposed pipelines over traditional STFT-based models in terms of robustness and subjective ratings. Additionally, the codec-based approach effectively reduces uncertainty and ambiguity in restoration mapping, showing notable advantages in real-world scenarios.

\section{Proposed Approach}
\label{sec:algorithm}



\subsection{Vocoder Approach}

As depicted in Fig. \ref{fig:vocoder} we illustrate the vocoder approach, wherein a noisy mel-spectrogram is transformed into a clean counterpart using an acoustic enhancer. During inference, we leverage a pre-trained HifiGAN vocoder \cite{kong2020hifigan} to restore the clean speech. 
An auxiliary input is produced by employing an SSL conditioner on the SSL features.
Specifically, we adopt the \texttt{WavLM-Large} variant of the WavLM model \cite{chen2022wavlm}, extract the learnable weighted sum of all layered results to produce 1024-dimension SSL features, which is then processed by the SSL conditioner to extract the SSL embedding of 256 dimensions.
This conditioner comprises a three-layer 1-dimensional convolutional network with upsampling, ReLU activation, instance normalization, and a dropout of 0.5.
The acoustic enhancer, based on deep complex convolutional recurrent network (DCCRN) architecture \cite{hu2020dccrn}, employs a convolutional encoder-decoder with an LSTM bottleneck.
Concretely, DCCRN consists of a six-layer convolution encoder and decoder, and an LSTM block in the bottleneck part to model time dependencies. We adjust the architecture for mel-spectrogram input by removing all complex-value related operations and setting the input convolutional channels to 1. 
The auxiliary input is fed to the bottleneck and is concatenated with the input of the LSTM block.
To make the training more efficient, the vocoder modules are only used during inference.
During training, we calculate the L1 loss between enhanced and clean mel-spectrograms.
Given the degraded speech input $x \in \mathbb{R}^L$, the target clean speech $y$ is of the same length $L$.
For the intermediate representation, we extract 128-band mel-spectrograms at a hop-length of 10 ms with a Hann window of 64 ms, resulting in mel features $Mel(x) \in \mathbb{R}^{T\times K}$, where $T$ denotes the number of time frames and $K$ represents the feature dimension 128. The training objective for the vocoder approach is then defined as,
\begin{equation}
    \mathcal{L}^{Vocoder} = \frac{1}{TK} \sum_{t=1}^T\sum_{k=1}^K |\hat{M}- Mel(y)|, \label{eq:mel}
\end{equation}
where $\hat{M}$ is the estimated mel-spectrogram, and $Mel(y)$ represents the ground truth mel-spectrogram. 

\subsection{Codec Approach}

\begin{table}[!tbp]

\centering
\vspace{2mm}

\caption{Objective scores of Comparison of All Pipelines}
\resizebox{0.94\linewidth}{!}{
\begin{tabular}{@{}lccccccc@{}}
\toprule
                 &  \multicolumn{3}{c}{Reverberant Only} && \multicolumn{3}{c}{Noisy + Reverberant} \\ \cmidrule{2-4} \cmidrule{6-8}         
                 & STOI  & PESQ    & DNS-MOS && STOI  & PESQ     & DNS-MOS  \\ \midrule
Unprocessed     & 0.663 &  1.648   & 2.859   && 0.624 &  1.511   &  2.635     \\ \hdashline[1pt/2pt]\hdashline[0pt/1pt]
Vocoder\_Best    & \textbf{0.870} & \textbf{2.472}   & 3.579   && \textbf{0.825} &  \textbf{2.121}   &  3.452     \\
Codec\_Best      & 0.835 & 2.102   & \textbf{3.718}   && 0.802 &  1.916   &  \textbf{3.641}   \\
STFT\_Based      & 0.787 &  1.948  & 3.133   && 0.751 &  1.801   &  3.024    \\  \bottomrule
\end{tabular}
}
\vspace{-2mm}
\label{tbl:summary}
\end{table}

\begin{figure}[!thb]
\centering
\includegraphics[width=0.7\linewidth]{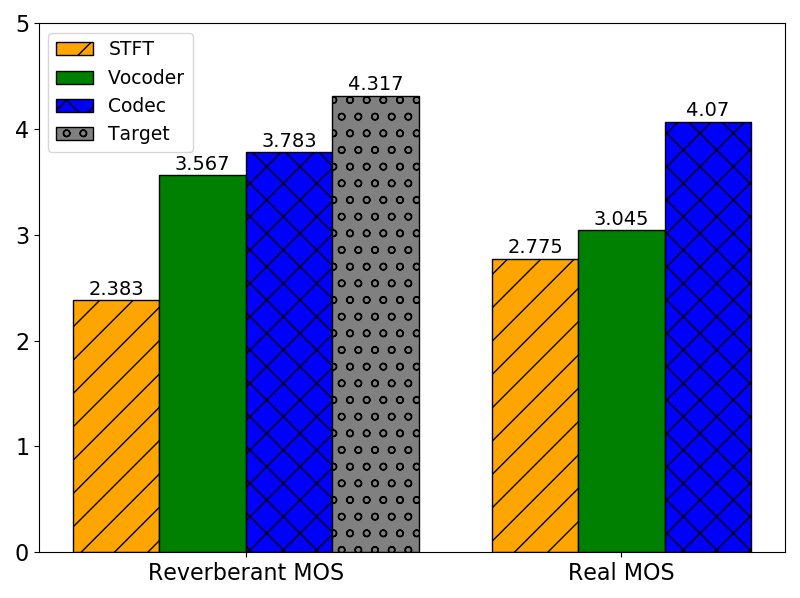}
\caption{Diagrams of MOS for both synthetic and realistic samples.}
\label{fig:mos}
\end{figure}

We depict the pipeline of the codec approach in Fig. \ref{fig:codec}.
The implementation entails the application of supervised enhancement learning within the code token space. 
This involves attempting to obtain the code tokens for the target speech and then using a pre-trained speech decoder to restore the clean speech.
The code enhancer is designed to predict clean code tokens based on the primary input codec embedding and the auxiliary input mel-spectrograms.
This undertaking is similar to a classification task focusing on code tokens.
Initial attempts to predict tokens corresponding to clean speech encountered challenges.
Firstly, most feature encoders of existing codecs are not trained using degraded speech utterances.
This inconsistency between the corrupted features of the codec input and the accurate derivation of code tokens by the codec led to instability in input code tokens, thereby yielding suboptimal enhancement outcomes.
Furthermore, predicting speech embeddings (either pre-vector quantization or post-vector quantization) is comparatively simpler. However, the generated speech by the decoder may contain distortions, as the predicted embeddings may not align well with the pre-stored patterns in the codebooks, consequently affecting enhancement performance.
To address these issues, we have proposed several techniques.
Firstly, a generalized codec architecture is adopted, involving a EnCodec \cite{defossez2022Encodec} trained on utterances from multiple languages retrieved from gigaspeech \cite{chen2021gigaspeech}, LibriTTS \cite{zen2019libritts}, and VP10K \cite{wang-etal-2021-voxpopuli} and common voice \cite{ardila2019common}, and augmented with 20\% probabilities by simulated noise and reverberations.
Additionally, encoder embeddings are employed as primary inputs during training, alongside mel-spectrograms of the input speech as auxiliary input.
Finally, we have worked on the training target and training objective.
Two techniques are investigated to mitigate prediction errors when the correct tokens cannot be retrieved.
The first is label smoothing. Specifically, principal component analysis is applied to the target tokens, and they are sorted based on the absolute values of the retrieved codebook entries. Label smoothing is then incorporated for neighboring tokens, where we set the target token with a probability of 0.9, and its neighbor tokens with 0.05.
For the other technique, we retrieve the codebook entries $Z$ from the tokens $C$ using the gumbel softmax layer \cite{jang2016gumbel} in a fully differentiable way,
which forms quantified representations of $Z_c \in \mathbb{R}^{N\times D}$, where $N$ to indicate the total number of code tokens and $D$ represents the feature dimension of each codebook entry.
It can be formulated as,
\vspace{-1.3mm}
\begin{align}
    \mathcal{L}^{Codec} &= \mathcal{L}^{Token} + \lambda \mathcal{L}^{Entry} \nonumber \\
    \mathcal{L}^{Token} &= \frac{1}{N} \sum_{i=1}^{N} C_i log(\hat{C_i})  \nonumber \\
    \mathcal{L}^{Entry} &= \frac{1}{ND} \sum_{i=1}^{N}\sum_{j=1}^{D} (Z_d - \hat{Z}_c)^2
\vspace{-2mm}
\end{align}
where $\mathcal{L}_{Token}$ measures the cross entropy loss of the predicted code tokens $\hat{C}$ and the target tokens $C$.
The second term $\mathcal{L}_{Entry}$ measures the mean squared loss value loss between the acquired codebook entries obtained by degraded speech $Z_d$ and the retrieved entries $Z_c$ acquired by the predicted tokens $\hat{C}$. 
The coefficient $\lambda$ is empirically chosen to be 0.5.

Given the importance of the first code index and the hierarchical architecture inherent in residual quantization, an architecture based on layer-wise modeling is adopted to enhance performance.
The proposed model architecture comprises a transformer decoder and a prediction layer. This transformer decoder integrates 12 transformer blocks, characterized by an embedding dimension of 512. The model input encompasses the codec embedding and the auxiliary input acoustic conditioner embedding. For the prediction of second to last code tokens, an additional embedding generated from preceding code tokens is incorporated. The prediction layer facilitates the projection of transformer outputs to 1024 dimensions, which corresponds to the size of the codebook vocabulary.

\subsection{Comparisons with Conventional DNN Approach}
\vspace{-2mm}
We compare the proposed two pipelines with the established DNN-based approach for speech enhancement, which directly operates on the speech representations, and maps noisy speech signals into their denoised counterparts.
Specifically, we adopt the DCCRN \cite{hu2020dccrn} as the STFT backbone.
This facilitates uniformity in our experimental setup, wherein we undertake speech enhancement trials employing identical datasets for the purpose of comparative analysis.

\vspace{-3mm}
\section{Experimental Setup}
\vspace{-2mm}
\label{sec:setup}
Our experiments are primarily conducted by performing dereverberation of LibriTTS utterances \cite{zen2019libritts}, where we extract clean utterances of 16k Hz with high quality as the target. The training subset comprises 147,039 instances of utterances, while the validation subset encompasses 5,566 distinct utterances. Subsequently, an evaluation is performed on 4,589 utterances that have not been encountered during training. 
For each utterance, we simulate the reverberation by convolving with a simulated room impulse response (RIR). 
The process of generating RIRs entails a stochastic selection of a $T_{60}$ parameter, denoting the reverberation duration, from an interval spanning 0.2 to 1.5 seconds, executed through the image method \cite{allen1979image}. 
In conjunction, the spatial dimensions of the room are determined by random selection, with width and length options ranging between 3, 4, and 2.5 meters, and height alternatives within 10, 20, and 4 meters. 
In addition, for the noisy-reverberant dataset simulation, we additionally add environmental noises with signal-to-noise-ratio (SNR) ranging from 0 to 40 dB. 
During training, an Adam optimizer is employed. The training process is instantiated with a batch size of 32 utterances, accompanied by an initial learning rate of 4e-4, which is sustained across 400 epochs.
During training, we randomly cut a 4-second segment for each training utterance, and pad shorter utterances with zeros within each batch to guarantee they are of the same size.

The performance assessment of the models is executed through a comparative analysis of the resynthesized clean speech against the reference dry clean speech. 
To this end, two established metrics are employed: the short-time objective intelligibility (STOI) \cite{taal2011stoi} metric and the perceptual evaluation of speech quality (PESQ) \cite{rix2001perceptual}.
In addition, we provide an evaluation of subjective metrics, which include the non-intrusive DNS-MOS \cite{reddy2022dnsmos} metric, and mean opinion scores (MOS) that are provided by human listeners. 
The MOS evaluation includes synthetic and real-world meeting data components. Synthetic samples consist of 10 reverberant instances, featuring input, enhanced samples from three methodologies, and ground truth samples. The second segment involves 10 real recorded utterances captured in diverse settings using different recording devices. These utterances are intentionally shuffled for each evaluation instance within this segment.

\vspace{-3mm}
\section{Results and Analysis}
\vspace{-2mm}

\subsection{Comparison of All Pipelines}
\vspace{-1mm}

Table \ref{tbl:summary} summarizes the best experimental results on two generative pipelines and the STFT-based pipeline. The vocoder approach performs best in terms of STOI and PESQ, while the codec approach attains the best DNS-MOS scores and effectively removes background and reverberation. A subjective auditory assessment is graphically depicted in Fig. \ref{fig:mos}. Both codec and vocoder models surpass the STFT approach in synthetic samples. 
The difference between STOI and PESQ scores primarily arises from codec and vocoder characteristics. The oracle codec has lower subjective scores (Table \ref{tbl:codec_input} and \ref{tbl:vocoder_inp}). Moreover, the high compression rate of the codec may introduce slight resynthesis misalignment, resulting in decreased objective scores.
In real-world meeting scenarios, the codec technique presents a distinct advantage over the other two pipelines due to its superior interference robustness. This is because the codec decoder retains only clear speech patterns, effectively removing noise or reverberation, and utilizing a previously acquired discrete codebook reduces ambiguity in speech recovery.

\vspace{-3mm}
\subsection{Evaluation of the Vocoder Approach}
\label{ssec:vocoder}
\begin{table}[!tbp]
\centering
\caption{Compare Different Inputs to the Vocoder Using DCCRN}
\vspace{-2mm}
\resizebox{0.83\linewidth}{!}{
\begin{tabular}{@{}lcccc@{}}
\toprule
                            & STOI & PESQ & DNS-MOS  \\ \midrule
Unprocessed                 & 0.663 & 1.648 & 2.857  \\ \hdashline[1pt/2pt]\hdashline[0pt/1pt]
Last Layer (SSL Only)           & 0.718 & 1.459 & 3.084  \\
WS (SSL Only)                   & 0.731 & 1.438 & 3.163  \\
Mel-spectrogram Only        & 0.865 & 2.372 & 3.525  \\ \hdashline[1pt/2pt]\hdashline[0pt/1pt]
Mel + WS                    & 0.869 & 2.450 & 3.573  \\ 
Mel + LWS                   & \textbf{0.870} & \textbf{2.472} & \textbf{3.579}  \\ \hdashline[1pt/2pt]\hdashline[0pt/1pt]
Vocoder Oracle              & 0.957 & 3.611 & 3.740  \\ \bottomrule
\label{tbl:vocoder_inp}
\end{tabular}
}
\vspace{-3mm}
\end{table}
As presented in Table \ref{tbl:vocoder_inp}, a comparative analysis of diverse inputs was conducted within the framework of the vocoder approach. 
The initial two rows of investigation aim to map the SSL embedding to the enhanced mel-spectrogram directly.
Inspection of the results reveals that only utilizing SSL embeddings produces unsatisfactory outcomes, as the SSL embeddings contain sufficient semantic information, but at the same time lose speaker and timbre information. This information loss impedes the restoration clean speech.
In addition, the weighted sum (WS) of multiple layered SSL embeddings shows noticeable improvement across all metrics.
When combined with the auxiliary input, the enhanced performance is considerably improved.
Especially when we use the learnable weighted sum (LWS) of SSL embedding as auxiliary input, and mel-spectrogram as the primary input, we obtain the best enhancement performance.
Lastly, for reference, we provide the upper bound of this approach, that is the last row, where the ground truth mel-spectrogram is employed for the synthesis of the target speech.

\begin{table}[!tbp]
\centering
\caption{Ablation Study on Vocoder}
\vspace{-2mm}

\resizebox{0.95\linewidth}{!}{
\begin{tabular}{@{}lcccc@{}}
\toprule
                                  & STOI & PESQ & DNS-MOS \\ \midrule
Baseline (Mel + LWS)               & \textbf{0.870} & \textbf{2.472} & \textbf{3.579}  \\ \hdashline[1pt/2pt]\hdashline[0pt/1pt]
Add Adapter (i)                       & 0.865 & 2.394 & 3.567  \\ 
No Bottleneck (ii)                   & 0.855 & 2.377 & 3.550  \\
SSL Token  (iii)                       & 0.852  & 2.231  & 3.457 \\ 
Use Transformer Instead of DCCRN (iv) & 0.867 & 2.368 & 3.802  \\ \bottomrule

\label{tbl:vocoder_ablation}
\vspace{-8mm}
\end{tabular}
}

\end{table}

Table \ref{tbl:vocoder_ablation} presents the results of the ablation study on the vocoder approach.
Multiple variations of the vocoder approach have been examined:
(i) instead of bottleneck concatenation, directly concatenate SSL embeddings with the mel-spectrogram along the feature dimension; (ii) substitute continuous SSL representations with discrete SSL tokens extracted by k-means; (iii) introduce the residual adapter \cite{otake2023wavlmadapter} to extract SSL representations; (iv) a Transformer architecture akin to the codec approach is employed as the acoustic enhancer.
From experimental results, we observe that the current design performs the best. Adding a residual adapter is not computationally efficient, and does not surpass the advantages conferred by the mere employment of LWS. Furthermore, (iv) facilitates a fair comparison with the codec approach.


\vspace{-3mm}
\subsection{Evaluation of the Codec Approach}
\label{ssec:codec}
\begin{table}[!tbp]
\centering
\caption{Compare Different Inputs to the codec}

\resizebox{0.91\linewidth}{!}{
\begin{tabular}{@{}lcccc@{}}
\toprule
                & STOI  & PESQ  & DNS-MOS    \\ \midrule
Unprocessed      & 0.663 & 1.648 & 2.859  \\ \hdashline[1pt/2pt]\hdashline[0pt/1pt]
Code tokens          & 0.745 & 1.662 & 3.628  \\
SSL Embeddings   & 0.765 & 1.419 & 3.682  \\ 
Codec Embeddings & \textbf{0.807} & \textbf{1.918} & \textbf{3.685}  \\
Mel Spectrograms & 0.669 & 1.272 & 3.496  \\ \hdashline[1pt/2pt]\hdashline[0pt/1pt]
 + Mel Spectrograms     & \textbf{0.835} & \textbf{2.102} & \textbf{3.718} \\
 + SSL Embeddings       & 0.828 & 2.037 & 3.727  \\ 
 Mel + SSL Spectrograms & 0.825 & 2.014 & 3.729   \\  \hdashline[1pt/2pt]\hdashline[0pt/1pt]
 Codec Oracle     & 0.904 & 2.764  & 3.689 &  \\  \bottomrule

\end{tabular}
}
\label{tbl:codec_input}
\end{table}

\begin{table}[!tbp]
\centering
\vspace{1mm}
\caption{Ablation Studies on Codec}
\resizebox{0.99\linewidth}{!}{
\begin{tabular}{@{}lcccc@{}}
\toprule
                    & STOI & PESQ & DNS-MOS \\ \midrule
Baseline (CE + Entry + Layerwise)  & \textbf{0.835} & \textbf{2.102} & 3.718 \\ \hdashline[1pt/2pt]\hdashline[0pt/1pt]
- Entry loss (i)  & 0.832 & 2.071 & 3.714 \\ 
Replace Entry Loss with Label Smoothing (ii) & 0.833 & 2.076 & \textbf{3.751} \\
Add Label Smoothing (iii) & 0.821 & 1.980 & 3.701 \\
Layer-wise Token Prediction (iv) & 0.797 & 1.960 & 3.705 \\ \bottomrule
\end{tabular}
}
\label{tbl:codec_ablation}
\vspace{-3mm}

\end{table}

A comparative analysis was conducted to assess the impact of various input features on code token prediction, and the results are listed in Table \ref{tbl:codec_input}.
The results show that embeddings as the principal input exhibit superior performance compared to alternative inputs, augmenting all measured metrics. SSL embeddings improve STOI and DNS-MOS score but result in suboptimal PESQ performance due to information loss inherent in SSL embeddings during their extraction. The application of code tokens or mel-spectrogram does not yield optimal outcomes, leading to unwanted artifacts in generated speech.
Adding auxiliary input that contains semantic or acoustic information can bring benefits to the enhancement performance. Both mel-spectrograms and SSL embeddings help, but the computational overhead associated with SSL embeddings is notably higher. Consequently, the proposed framework adopts the mel-spectrogram as the preferred input. The result of the upperbound performance is also provided for reference.

Table \ref{tbl:codec_ablation} reports the results of ablation studies pertinent to the codec approach.
We use the best configuration, featuring layer-wise prediction and the utilization of cross entropy (CE) + Entry loss as the baseline, and compare several variants on the reverberant dataset:
(i) solely train using the CE loss for code tokens; (ii) substitute the entry loss with label smoothing; (iii) use these two techniques simultaneously; (iv) instead of predicting code tokens in a layer-wise manner, predict all tokens simultaneously.
As shown in the table, both label smoothing and adding entry loss are beneficial when the code tokens predictions are not accurate.
However, using these two techniques simultaneously produces sub-optimal results. Therefore, a solitary technique suffices.
Predicting all code tokens simultaneously degrades the performance, as it does not address the importance of the first code, and could not leverage teacher-forcing during training.

\vspace{-2mm}
\section{Conclusion}
\vspace{-1mm}

\label{sec:conclusion}
In conclusion, this study introduces an innovative approach that leverages pre-trained generative methods to address the long-standing challenges of enhancing speech signal quality in adverse acoustic environments.
By employing established vocoder, codec, and self-supervised learning models, the proposed methodology effectively resynthesizes clean and anechoic speech from degraded inputs, mitigating issues like background noise and reverberation. Through empirical evaluations in both simulated and real-world scenarios, the method demonstrates superior subjective scores, showcasing its ability to improve audio fidelity, reduce artifacts, and superior robustness. 
This research highlights the potential of leveraging generative techniques in speech processing, especially in challenging scenarios where conventional methods fall short.

\vfill\pagebreak
\clearpage

\bibliographystyle{IEEEbib}
\bibliography{refs}

\begin{thebibliography}{10}

\bibitem{wang2018supervised}
D.~L. Wang and J.~Chen,
\newblock ``Supervised speech separation based on deep learning: An overview,''
\newblock {\em IEEE/ACM Transactions on Audio, Speech, and Language
  Processing}, vol. 26, pp. 1702--1726, 2018.

\bibitem{han2014tms}
K.~Han, Y.~Wang, and D.~L. Wang,
\newblock ``Learning spectral mapping for speech dereverberation,''
\newblock in {\em Proceedings of ICASSP}, 2014, pp. 4628--4632.

\bibitem{li2020online}
X.~Li and R.~Horaud,
\newblock ``Online monaural speech enhancement using delayed subband {LSTM},''
\newblock in {\em Proceedings of INTERSPEECH}, 2020, pp. 2462--2466.

\bibitem{choi2018phase}
H.-S. Choi, J.-H. Kim, J.~H., A.~Kim, J.-W. Ha, and K.~Lee,
\newblock ``Phase-aware speech enhancement with deep complex {U-Net},''
\newblock in {\em Proceedings of ICLR}, 2018.

\bibitem{hu2020dccrn}
Y.~Hu, Y.~Liu, S.~Lv, M.~Xing, S.~Zhang, Y.~Fu, J.~Wu, B.~Zhang, and L.~Xie,
\newblock ``{DCCRN}: Deep complex convolution recurrent network for phase-aware
  speech enhancement,''
\newblock {\em arXiv:2008.00264}, 2020.

\bibitem{luo2018tasnet}
Y.~Luo and N.~Mesgarani,
\newblock ``Tasnet: time-domain audio separation network for real-time,
  single-channel speech separation,''
\newblock in {\em Proceedings of ICASSP}, 2018, pp. 696--700.

\bibitem{pandey2021dense}
A.~Pandey and D~L Wang,
\newblock ``Dense {CNN} with self-attention for time-domain speech
  enhancement,''
\newblock {\em IEEE/ACM Transactions on Audio, Speech, and Language
  Processing}, vol. 29, pp. 1270--1279, 2021.

\bibitem{li2021two}
A.~Li, W.~Liu, C.~Zheng, C.~Fan, and X.~Li,
\newblock ``Two heads are better than one: A two-stage complex spectral mapping
  approach for monaural speech enhancement,''
\newblock {\em IEEE/ACM Transactions on Audio, Speech, and Language
  Processing}, vol. 29, pp. 1829--1843, 2021.

\bibitem{wang2021nca}
H.~Wang and D.~L. Wang,
\newblock ``Neural cascade architecture with triple-domain loss for speech
  enhancement,''
\newblock {\em IEEE/ACM Transactions on Audio, Speech, and Language
  Processing}, vol. 30, pp. 734--743, 2021.

\bibitem{rao2021interspeech}
W.~Rao, Y.~Fu, Y.~Hu, X.~Xu, Y.~Jv, J.~Han, Z.~Jiang, L.~Xie, Y.~Wang,
  S.~Watanabe, et~al.,
\newblock ``{INTERSPEECH} 2021 conferencing speech challenge: Towards far-field
  multi-channel speech enhancement for video conferencing,''
\newblock {\em arXiv:2104.00960}, 2021.

\bibitem{wang2023cross}
H.~Wang and D.~L. Wang,
\newblock ``Cross-domain diffusion based speech enhancement for very noisy
  speech,''
\newblock in {\em Proceedings of ICASSP}, 2023, pp. 1--5.

\bibitem{lemercier2023storm}
J.-M. Lemercier, J.~Richter, S.~Welker, and T.~Gerkmann,
\newblock ``{StoRM}: A diffusion-based stochastic regeneration model for speech
  enhancement and dereverberation,''
\newblock {\em IEEE/ACM Transactions on Audio, Speech, and Language
  Processing}, 2023.

\bibitem{liu2022voicefixer}
H.~Liu, X.~Liu, Q.~Kong, Q.~Tian, Y.~Zhao, D.~L. Wang, C.~Huang, and Y.~Wang,
\newblock ``Voicefixer: A unified framework for high-fidelity speech
  restoration,''
\newblock {\em arXiv:2204.05841}, 2022.

\bibitem{kong2020hifigan}
J.~Kong, J.~Kim, and J.~Bae,
\newblock ``{HiFi-GAN}: Generative adversarial networks for efficient and high
  fidelity speech synthesis,''
\newblock {\em Advances in Neural Information Processing Systems}, vol. 33, pp.
  17022--17033, 2020.

\bibitem{zhong2023audiomae}
Zhi Zhong, Hao Shi, Masato Hirano, Kazuki Shimada, Kazuya Tateishi, Takashi
  Shibuya, Shusuke Takahashi, and Yuki Mitsufuji,
\newblock ``Extending audio masked autoencoders toward audio restoration,''
\newblock {\em arXiv:2305.06701}, 2023.

\bibitem{borsos2023audiolm}
Z.~Borsos, R.~Marinier, D.~Vincent, E.~Kharitonov, O.~Pietquin, M.~Sharifi,
  D.~Roblek, O.~Teboul, D.~Grangier, M.~Tagliasacchi, et~al.,
\newblock ``{AudioLM}: a language modeling approach to audio generation,''
\newblock {\em IEEE/ACM Transactions on Audio, Speech, and Language
  Processing}, 2023.

\bibitem{wang2023valle}
C.~Wang, S.~Chen, Y.~Wu, Z.~Zhang, L.~Zhou, S.~Liu, Z.~Chen, Y.~Liu, H.~Wang,
  J.~Li, et~al.,
\newblock ``Neural codec language models are zero-shot text to speech
  synthesizers,''
\newblock {\em arXiv:2301.02111}, 2023.

\bibitem{zhou2022codeformer}
S.~Zhou, K.~Chan, C.~Li, and C.C. Loy,
\newblock ``Towards robust blind face restoration with codebook lookup
  transformer,''
\newblock {\em Advances in Neural Information Processing Systems}, vol. 35, pp.
  30599--30611, 2022.

\bibitem{hu2023wav2code}
Y.~Hu, C.~Chen, Q.~Zhu, and E.S. Chng,
\newblock ``Wav2code: Restore clean speech representations via codebook lookup
  for noise-robust {ASR},''
\newblock {\em arXiv:2304.04974}, 2023.

\bibitem{chen2022wavlm}
S.~Chen, C.~Wang, Z.~Chen, Y.~Wu, S.~Liu, Z.~Chen, J.~Li, N.~Kanda,
  T.~Yoshioka, X.~Xiao, et~al.,
\newblock ``{WavLM}: Large-scale self-supervised pre-training for full stack
  speech processing,''
\newblock {\em IEEE Journal of Selected Topics in Signal Processing}, vol. 16,
  pp. 1505--1518, 2022.

\bibitem{defossez2022Encodec}
A.~D{\'e}fossez, J.~Copet, G.~Synnaeve, and Y.~Adi,
\newblock ``High fidelity neural audio compression,''
\newblock {\em arXiv:2210.13438}, 2022.

\bibitem{chen2021gigaspeech}
G.~Chen, S.~Chai, G.~Wang, J.~Du, W.-Q. Zhang, C.~Weng, D.~Su, D.~Povey,
  J.~Trmal, J.~Zhang, et~al.,
\newblock ``Gigaspeech: An evolving, multi-domain asr corpus with 10,000 hours
  of transcribed audio,''
\newblock {\em arXiv:2106.06909}, 2021.

\bibitem{zen2019libritts}
H.~Zen, V.~Dang, R.~Clark, Y.~Zhang, R.~J. Weiss, Y.~Jia, Z.~Chen, and Y.~Wu,
\newblock ``{LibriTTS}: A corpus derived from {LibriSpeech} for
  text-to-speech,''
\newblock {\em arXiv:1904.02882}, 2019.

\bibitem{wang-etal-2021-voxpopuli}
C.~Wang, M.~Riviere, A.~Lee, A.~Wu, C.~Talnikar, D.~Haziza, M.~Williamson,
  J.~Pino, and E.~Dupoux,
\newblock ``{V}ox{P}opuli: A large-scale multilingual speech corpus for
  representation learning, semi-supervised learning and interpretation,''
\newblock in {\em Proceedings of ACL}, 2021, pp. 993--1003.

\bibitem{ardila2019common}
R.~Ardila, M.~Branson, K.~Davis, M.~Henretty, M.~Kohler, J.~Meyer, R.~Morais,
  L.~Saunders, F.M. Tyers, and G.~Weber,
\newblock ``Common voice: A massively-multilingual speech corpus,''
\newblock {\em arXiv:1912.06670}, 2019.

\bibitem{jang2016gumbel}
E.~Jang, S.~Gu, and B.~Poole,
\newblock ``Categorical reparameterization with gumbel-softmax,''
\newblock {\em arXiv:1611.01144}, 2016.

\bibitem{allen1979image}
J.B Allen and D.A Berkley,
\newblock ``Image method for efficiently simulating small-room acoustics,''
\newblock {\em The Journal of the Acoustical Society of America}, vol. 65, pp.
  943--950, 1979.

\bibitem{taal2011stoi}
C.~H. Taal, R.~C. Hendriks, R.~Heusdens, and J.~Jensen,
\newblock ``An algorithm for intelligibility prediction of time--frequency
  weighted noisy speech,''
\newblock {\em IEEE/ACM Transactions on Audio, Speech, and Language
  Processing}, vol. 19, pp. 2125--2136, 2011.

\bibitem{rix2001perceptual}
A.~W. Rix, J.~G. Beerends, M.~P. Hollier, and A.~P. Hekstra,
\newblock ``Perceptual evaluation of speech quality ({PESQ})-a new method for
  speech quality assessment of telephone networks and codecs,''
\newblock in {\em Proceedings of ICASSP}, 2001, pp. 749--752.

\bibitem{reddy2022dnsmos}
C.~KA. Reddy, V.~Gopal, and R.~Cutler,
\newblock ``{DNSMOS P. 835}: A non-intrusive perceptual objective speech
  quality metric to evaluate noise suppressors,''
\newblock in {\em Proceedings of ICASSP}, 2022, pp. 886--890.

\bibitem{otake2023wavlmadapter}
Shinta Otake, Rei Kawakami, and Nakamasa Inoue,
\newblock ``Parameter efficient transfer learning for various speech processing
  tasks,''
\newblock in {\em Proceedings of ICASSP}, 2023, pp. 1--5.

\end{thebibliography}

\end{document}